\documentclass[11pt]{article}

\usepackage[includeheadfoot,a4paper,textwidth=170mm,head=6mm,
            headsep=8mm,top=10mm,foot=12mm,bottom=10mm]{geometry}
\usepackage{booktabs} 
\usepackage{amssymb,latexsym,amsfonts,amsmath}
\usepackage{graphicx}
\usepackage{enumerate}
\usepackage{multirow}
\usepackage{mathtools}
\usepackage{subfigure}
\newcommand{\pre}[1]{\prescript{#1}{}}
\usepackage[export]{adjustbox}
\usepackage{framed}
\usepackage{tcolorbox}
\usepackage{chngcntr}

\usepackage{wrapfig}
\usepackage{diagbox}

\usepackage[normalem]{ulem}

\usepackage{tikz}
\usetikzlibrary{calc,shapes,arrows}
\usepackage{scalerel}
\usepackage{enumitem}

\usepackage{url}
\usepackage{breakurl}
\usepackage[breaklinks]{hyperref}

\usepackage{tikz}
\usepackage{bm}

\usepackage{dsfont}
\usepackage[bb=boondox]{mathalfa}
\usepackage{lipsum}

\newtheorem{theorem}{Theorem}[section]

\newtheorem{definition}[theorem]{Definition}

\newtheorem{assumption}[theorem]{Assumption}
\numberwithin{equation}{section}

\newcommand{\II}{\mathsf{id}}

\newcommand{\R}{{\mathbb{R}}}

\newcommand{\TF}{\mathcal{F}}
\newcommand{\x}{\mathbf{x}}

\newcommand{\N}{{\mathbb{N}}}

\newcommand{\op}{{\mathcal H}}

\newcommand{\SF}{\mathcal{S}}

\newcommand{\n}{\mathcal{N}}
\newcommand{\ca}{\mathcal{C}}

\newcommand{\Let}{:=}

\def\sc#1{{\scaleto{#1}{6pt}}}

\definecolor{myco}{rgb}{0.2660 0.5740 0.1980}

\newcommand{\Z}{\mathbb{Z}}

\newcommand{\intcc}[1]{\ensuremath{{\left[#1\right]}}}

\title{{\LARGE \bf Finite Abstractions of Network of Impulsive Systems Using Dissipativity Approach}}
\author{ Abdalla Swikir 
}
\date{}
\onecolumn

\begin{document}
\maketitle

\begin{abstract}
This paper introduces a compositional framework for constructing finite abstractions of nonlinear interconnected impulsive systems using dissipativity-based conditions. Central to our approach is the concept of "alternating simulation functions," which serve to relate the concrete dynamics of impulsive subsystems to their finite  abstractions. Dissipativity conditions are employed to ensure the compositionality of the finite abstractions, enabling the representation of complex interconnected systems as compositions of their subsystem abstractions. The methodology relies on incremental passivity properties such as supply rates and storage functions, alongside forward completeness, to construct finite abstractions for individual impulsive subsystems.
\end{abstract}

\section{Introduction}

Finite abstractions(a.k.s.symbolic models) provide a powerful method for simplifying complex dynamical systems by representing them through finite sets of states, inputs, and transitions that capture the core dynamics of the original system. Formal relationships, such as simulation or alternating simulation \cite{cassandras2009introduction}, link these abstract models to the concrete systems, enabling tasks like model checking and controller synthesis. These techniques are particularly valuable for designing controllers that adhere to high-level specifications expressed in temporal logic \cite{tabuada2009verification}. However, the need for state and input space discretization often results in exponential computational complexity, making the abstraction process computationally expensive, especially for systems with high-dimensional state spaces. This phenomenon, commonly referred to as the "curse of dimensionality," poses significant challenges.

To address the challenges posed by large-scale interconnected systems, compositional abstraction has emerged as a promising approach. By decomposing the abstraction process into subsystem-level constructions, compositional methods enable a more scalable and efficient handling of interconnected systems. Significant progress has been made in this area, leading to frameworks for the abstraction of acyclic interconnected linear \cite{lal2019compositional}, nonlinear \cite{saoud2018compositional}, and discrete-time time-delay systems \cite{9304184}; methods based on dissipativity properties \cite{zamani2017compositional}; and abstraction frameworks for interconnected switched systems \cite{8796176,swikir2019compositionalLcss}.

The compositional abstraction of interconnected discrete-time control systems is addressed in \cite{8307949}, where a systematic framework is developed for constructing symbolic models. Similarly, \cite{8550268} investigates the compositional construction of approximately bisimilar abstractions for incrementally input-to-state stable networks of systems. In addition, compositional synthesis of abstractions for infinite networks has been explored in \cite{SWIKIR20201868,SHARIFI2022101173,9304074,LIU2021101097}. Further approaches include assume-guarantee contracts \cite{saoud2020contract,saoud2021assume} and approximate composition methods \cite{saoud2021compositional,sadek2022compositional}. A comprehensive discussion of compositional frameworks can be found in \cite{dissertation}.

Despite these advancements, compositional construction of finite abstractions of impulsive systems remains underexplored. Prior work, such as \cite{swikir2020symbolic}, addresses the monolithic abstraction of impulsive systems, which becomes computationally burdensome for large-scale interconnected systems. A notable recent development in \cite{10383255} presents a compositional methodology for constructing symbolic models of nonlinear interconnected impulsive systems using small-gain-type conditions. However, the reliance on gain constraints limits the applicability of this approach, particularly for certain interconnection topologies.

In this paper, we address these limitations by proposing a novel compositional framework for constructing finite abstractions of a network of impulsive systems based on dissipativity-type conditions. Unlike small-gain-type approaches, dissipativity-based conditions eliminate the need for gain constraints in some interconnection topologies, significantly broadening the applicability of the framework. Our methodology extends the concept of alternating approximate simulation functions \cite{pola2009symbolic} to establish formal relationships between subsystems and their symbolic models. By leveraging dissipativity properties and forward completeness, we develop a systematic process for constructing finite abstractions for individual subsystems and for the overall interconnected system.

\section{Notations and Preliminaries}\label{1:II}
\paragraph*{\textbf{Notations:}}\label{note}
We denote by $\R$, $\Z$, and $\N$ the set of real numbers, integers, and non-negative integers,  respectively.
These symbols are annotated with subscripts to restrict them in
an obvious way, e.g., $\R_{>0}$ denotes the positive real numbers. We denote the closed, open, and half-open intervals in $\R$ by $[a,b]$,
$(a,b)$, $[a,b)$, and $(a,b]$, respectively. For $a,b\in\N$ and $a\leqslant b$, we
use $[a;b]$, $(a;b)$, $[a;b)$, and $(a;b]$ to
denote the corresponding intervals in $\N$.
Given any $a\in\R$, $| a|$ denotes the absolute value of $a$. Given any $u=[u_1;\ldots;u_n]\in\R^{n}$, the infinity norm of $u$ is defined by $\| u\|=\max_{i\in[1;n]}\|u_i\|$.  
Given a function $\nu: \R_{\ge0} \rightarrow \R^n $, the supremum of $ \nu $ is denoted by $ \| \nu\|_\infty $; we recall that $ \| \nu\|_\infty := \text{sup}_{t\in\R_{\ge0}}\| \nu(t)\|$. Given $\mathbf x:\R_{\geqslant0}\rightarrow\R^{n},\forall t, s\in\R_{\geqslant0}$ with $t\geqslant s$, we define $\mathbf{x}(\pre{-}t)=\lim_{s\rightarrow t}\mathbf{x}(s)$ as the left limit operator. For a given constant $\tau\in\R_{\ge0}$ and a set  $\mathcal{W}:=\{\mathbf x:\R_{\geqslant0}\rightarrow\R^{n}\}$, we denote the restriction of $\mathcal{W}$ to the interval $[0,\tau]$ by $\mathcal{W}|_{[0,\tau]}:=\{\mathbf x:[0,\tau]\rightarrow\R^{n}\}$.
We denote by $\ca(\cdot)$ the cardinality of a given set and by $\varnothing$ the empty set. Given sets $U$ and $S\subset U$, the complement of $S$ with respect to $U$ is defined as $U\backslash S = \{x : x \in U, x \notin S\}$. Given a family of finite or countable sets $S_i, i\in\n\subset\N$, the $j^{th}$ element of the set $S_i$ is denoted by $s_{i_j}$. For any set $S\subseteq\R^n$ of the form $S=\bigcup_{j=1}^MS_j$ for some $M\in \N_{>0}$, where $S_j=\prod_{i=1}^n [c_i^j,d_i^j]\subseteq \R^n$ with $c^j_i<d^j_i$, and non-negative constant $\eta\leqslant\tilde{\eta}$, where $\tilde{\eta}=\min_{j=1,\ldots,M}\eta_{S_j}$ and \mbox{$\eta_{S_j}=\min\{|d_1^j-c_1^j|,\ldots,|d_n^j-c_n^j|\}$}, we define \mbox{$[S]_{\eta}=\{a\in S\,\,|\,\,a_{i}=k_{i}\eta,k_{i}\in\mathbb{Z},i=1,\ldots,n\}$} if $\eta\neq0$, and $[S]_{\eta}=S$ if $\eta=0$. The set $[S]_{\eta}$ will be used as a finite approximation of the set $S$ with precision $\eta\neq0$. Note that $[S]_{\eta}\neq\varnothing$ for any $\eta\leqslant\tilde{\eta}$. 
We use notations $\mathcal{K}$ and $\mathcal{K}_\infty$
to denote different classes of comparison functions, as follows:
$\mathcal{K}=\{\alpha:\mathbb{R}_{\geqslant 0} \rightarrow \mathbb{R}_{\geqslant 0} |$ $ \alpha$ is continuous, strictly increasing, and $\alpha(0)=0\}$; $\mathcal{K}_\infty=\{\alpha \in \mathcal{K} |$ $ \lim\limits_{s \rightarrow \infty} \alpha(s)=\infty\}$.
For $\alpha,\gamma \in \mathcal{K}_{\infty}$ we write $\alpha\le\gamma$ if $\alpha(r)\le\gamma(r)$, $\forall r\in\R_{\geqslant0}$, and, by abuse of notation, $\alpha=c$ if $\alpha(r)=cr$ for all $c,r\geqslant0$. Finally, we denote by $\II$ the identity function over $\R_{\ge0}$, i.e. $\II(r)=r, \forall r\in \R_{\ge0}$.

\subsection{Concrete System} 

\subsubsection{Impulsive Nonlinear Subsystems} 

We consider a collection of impulsive subsystems indexed by $i \in \n$, where $\n = [1;N]$ and $N \in \mathbb{N}_{\geqslant 1}$. Each subsystem, indexed by $i$, is formally defined as follows: 
\begin{definition}\label{def:sys1}
A nonlinear impulsive subsystem $\Sigma_i$, $i \in \mathcal{N}$, is represented by the tuple \\$\Sigma_i=(\mathbb{R}_{i}^{n_{i}},\mathbb W_i,\mathsf{W}_i,\mathbb U_i,\mathsf{U}_i,f_i,g_i)$,
where 
\begin{itemize}
\item $\R_{i}^{n_i}$ is the state set; 
\item $\mathbb W_i\subseteq\R^{q_i}$ is the internal input set; 
\item $\mathsf{W}_i$ is the set
of all measurable bounded internal input functions $\omega_i:\R_{\geqslant0}\rightarrow \mathbb W_i$;
\item $\mathbb U_i\subseteq\R^{m_i}$ is the external input set;
\item $\mathsf{U}_i$ is the set
of all measurable bounded external input functions $\nu_i:\R_{\geqslant0}\rightarrow \mathbb U_i$;
\item $f_i,g_i: \R^{n_i}\times \mathbb W_i\times \mathbb U_i \rightarrow \R^{n_i} $ are locally Lipschitz functions;
\item $\Omega_i=\{t_i^k\}_{k\in\N}$ is a set of strictly increasing sequence of impulsive times in $\R_{\ge0}$ comes with $t_i^{k+1}-t_i^{k}\in\{\underline{z}_i\tau_i,\ldots,\overline{z}_i\tau_i\}$ for fixed jump parameters $\tau_i\in\R_{>0}$ and $\underline{z}_i,\overline{z}_i\in \N_{\ge1}$, $\underline{z}_i\le \overline{z}_i$.
\end{itemize}

The nonlinear flow and jump dynamics, $f_i$ and $g_i$, are characterized by the following differential and difference equations:
\begin{align}\label{eq:2}
\Sigma_i:\begin{cases}
		\mathbf{\dot{x}}_i(t)= f_i(\mathbf{x}_i(t),\omega_i(t),\nu_i(t)), & t\in\R_{\geqslant0}\backslash \Omega_{i},\\
		\mathbf{x}_i(t)=g_i(\mathbf{x}_i(\pre{-}t),\omega_i(\pre{-}t),\nu_i(t)),& t\in \Omega_i,\\
	\end{cases} 
\end{align} 
where $\mathbf{x}_i:\R_{\geqslant0}\rightarrow \R^{n_i}$ and  $\omega_i:\R_{\geqslant0}\rightarrow \mathbb W_i$ are the state and internal input signals, respectively, and assumed to be right-continuous for all $t\in\R_{\geqslant0}$, and function $\nu_i :\R_{\geqslant0}\rightarrow \mathbb U_i$, is the external input signal. We will use $\mathbf{x}_{x_i,\omega_i,\nu_i}(t)$ to denote a point reached at time $t\in \R_{\geqslant0}$ from initial state $x_i$ under input signals $\omega_i\in\mathsf{W}_i$ and $\nu_i\in\mathsf{U}_i$. We denote by $\Sigma_{c_i}$ and $\Sigma_{d_i}$ the continuous and discrete dynamics of subsystem $\Sigma_i$, i.e., $\Sigma_{c_i}:\mathbf{\dot{x}}_i(t)= f_i(\mathbf{x}_i(t),\omega_i(t),\nu_i(t))$, and $\Sigma_{d_i}:\mathbf{x}_i(t)= g_i(\mathbf{x}_i(\pre{-}t),\omega_i(\pre{-}t),\nu_i(t))$. 
\end{definition}
\subsubsection{Network of Impulsive Systems}
The formal definition of the network of impulsive systems is expressed as:
\begin{definition}
\label{interconnectedsystem} 
Consider $N \in \mathbb{N}_{\geqslant 1}$ impulsive subsystems, $\Sigma_i=(\R^{n_i},\mathbb W_i,\mathsf{W}_i,\mathbb U_i,\mathsf{U}_i,f_i,g_i)$ and a matrix $M$ of an appropriate dimension defining the coupling of these subsystems.
The network of impulsive systems is a tuple $\Sigma = (\mathbb X,\mathbb U,f,\mathcal{G},\Omega)$, denoted by $\mathcal{I}(\Sigma_1,\ldots,\Sigma_N)$ and described by the differential, difference equation of the form,
\begin{align}\label{intdyn}
 \Sigma : \begin{cases}
\dot{\mathbf{x}}(t) = f(\mathbf{x}(t),\nu(t)), & \forall t\in \R_{\geqslant 0}\backslash\Omega \\
\mathbf{x}(t)=\mathcal{G}(\mathbf{x}(\pre{-}t),\nu(t)) & \forall t\in \Omega
\end{cases}
\end{align}
with $x\in \!\mathbb X\!=\!\prod_{i=1}^N \R^{n_i}$, $\nu\! \in \!\mathbb{U}\!=\!\prod_{i=1}^N \mathbb{U}_i$, $\Omega\!=\!\bigcup_{i=1}^N\Omega_{i}$ and 
\begin{align*}
  f(\mathbf{x}(t),\nu(t))= \left[f_{1}(x_1(t),\omega_1(t),v_1(t)), \dots, f_n(x_n(t),\omega_n(t),v_n(t))\right] \\
\mathcal{G}(\mathbf{x}(\pre{-}t),\nu(t))= \left[\beta_{1}(x_1(\pre{-}t),\omega_1(\pre{-}t),v_1(t)), \dots, \beta_n(x_n(\pre{-}t),\omega_n(\pre{-}t),v_n(t))\right]
\end{align*}
where, 
\begin{align*}
\beta_i(x_i(\pre{-}t),\omega_i(\pre{-}t),v_i(t)) =\begin{cases}
    x_i(\pre{-}t) & if \; t \notin \Omega_i \\
    g_i(x_i(\pre{-}t),\omega_i(\pre{-}t),v_i(t)) & if \; t \in \Omega_i
\end{cases}
\end{align*}
and with the internal variables
constrained by\begin{align}
\label{internalinput}	
 	\left[\omega_{1} ; \ldots ; \omega_{N}\right]=M\left[\x_{1}\ ; \ldots ; x_{N}\right] .
  \end{align}
\end{definition}

\subsection{Transition systems}

\subsubsection{Transition Subsystems}\label{I}
Now, we will introduce the class of transition subsystems \cite{Tabu}, which will be later interconnected to form an interconnected transition system. Indeed, the concept of transition subsystems permits to model impulsive subsystems and
their finite abstractions in a common framework.
	\begin{definition}\label{ts} A transition subsystem is a tuple $T_i=(X_i,X_{0_i},W_i,\mathcal{W}_i,U_i,\mathcal{U}_i,\TF_i)$, $i\in\mathcal{N}$, consisting of:
		\begin{itemize}
			\item a set of states $X_i$;
		    \item a set of initial states $X_{0_i}\subseteq X_i$;
			\item a set of internal inputs values $W_i$;
			\item a set of internal inputs signals $\mathcal{W}_i\! \! :=\!\! \{\omega_i:\R_{\geq0}\rightarrow W_i\}$;
			\item a set of external inputs values $U_i$;
			\item a set of external inputs signals $\mathcal{U}_i:=\{u_i:\R_{\geq0}\rightarrow U_i\}$;
			\item transition function $\TF_i: X_i\times \mathcal{W}_i\times \mathcal{U}_i \rightrightarrows  X_i$.
		\end{itemize}
	\end{definition}
	The transition $x_i^+ \in \TF_i(x_i, \omega_i, u_i)$ signifies that the system can transition from state $x_i$ to state $x_i^+$ under the influence of the input signals $\omega_i$ (internal) and $u_i$ (external). Thus, the transition function $\TF_i$ encapsulates the dynamics governing the state transitions of the system. Let $\mathsf{x}_{x_i, \omega_i, u_i}$ denote an infinite state trajectory (or run) of the transition system $T_i$, associated with the initial state $x_i$, internal input signal $\omega_i$, and external input signal $u_i$.

The sets $X_i$, $\mathcal{W}_i$, and $\mathcal{U}_i$ are assumed to be subsets of normed vector spaces of appropriate finite dimensions. If, for all $x_i \in X_i$, $\omega_i \in \mathcal{W}_i$, and $u_i \in \mathcal{U}_i$, the cardinality of the transition function satisfies $\ca(\TF_i(x_i, \omega_i, u_i)) \leq 1$, then $T_i$ is said to be deterministic; otherwise, it is non-deterministic.

Additionally, $T_i$ is termed finite if $X_i$, $\mathcal{W}_i$, and $\mathcal{U}_i$ are finite sets; otherwise, $T_i$ is considered infinite. Furthermore, $T_i$ is called non-blocking if, for every $x_i \in X_i$, there exist $\omega_i \in \mathcal{W}_i$ and $u_i \in \mathcal{U}_i$ such that $\ca(\TF_i(x_i, \omega_i, u_i)) \neq 0$.

\subsubsection{Network of transition system}
We define the composed transition system by $ \mathcal{I}(T_1,\ldots,T_N) $ and we define it formally as
\begin{definition}
\label{tinterconnectedsystem} 
Consider $N \in \N_{\geqslant 1}$ transition subsystems $T_i=(X_i,X_{0_i},W_i,\mathcal{W}_i,U_i,\mathcal{U}_i,\TF_i)$. Let $x = [x_1;\dots;x_N]\in X$, $u = [u_1;\dots;u_N]\in U$, and $\|\left[\omega_{1} ; \ldots ; \omega_{N}\right]-M\left[\x_{1}\ ; \ldots ; x_{N}\right] \|\leqslant \left[\Phi_{1} ; \ldots ; \Phi_{N}\right], \quad \Phi_{i} \in \mathbb{R}_{\geq 0}$.
The interconnected transition system is a tuple $T=(X,X_{0},U,\TF)$, denoted by $\mathcal{I}(T_1,\ldots,T_N)$, where $X=\prod_{i=1}^N X_i$, $X_0=\prod_{i=1}^N X_{0_i}$, $ U=\prod_{i=1}^N U_i$.  Moreover, the transition relation $\TF$ is defined by,
\begin{align}
\label{eqn:transition}
\TF(x,u)&\!\Let\!\{\intcc{x^+_1;\ldots;x^+_N}\,|\, \!x^+_i\in \!\TF_i(x_i,u_i,\omega_i)~ \forall i\!\in\! \n \},
\end{align}

\end{definition}

\subsection{Alternating Simulation Function}

In this section, we recall the so-called notion of $\varepsilon-$ approximate alternating simulation function in \cite{swikir2019compositional}. 

\begin{definition}\label{sf} 
Let $T=(X,X_{0},U,\TF)$ and $\hat T=(\hat{X},\hat{X}_{0},\hat{U},\hat{\TF})$ with $\hat{X}\subseteq X$. A function $ \tilde{\mathcal{S}}:X\times \hat{X} \to \mathbb{R}_{\geqslant0} $ is called an alternating simulation function from $ \hat{T}$ to $\hat T$
if there exist $\tilde{\alpha} \in \mathcal{K}_{\infty}$,  $0<\tilde{\sigma}< 1$, $ \tilde{\rho}_{u} $ $\in \mathcal{K}_{\infty}\cup \{0\} $, and some $\tilde{\varepsilon}\in \mathbb{R}_{\geqslant 0}$ so that the following hold:
\begin{enumerate}
\item For every $ x\in X,\hat{x}\in\hat{X}$, we have,
\begin{align}\label{sf1}
	\tilde{\alpha} (\| x-\hat{x}\|) \!\leqslant\! \tilde{\SF}(x,\hat{x});
\end{align}
\item For every $x\in X,\hat{x}\in\hat{X},\hat u\in\hat{U}$ there exists $u\in U$ such that for every $ x^+\in\TF(x,u)$  there exists $\hat{x}^+\in\hat{\TF}(\hat{x},\hat{u})$ so that,
\begin{align}\label{sf2}
	\tilde{\SF}&(x^+,\hat{x}^+)\leqslant \max\{\tilde{\sigma} \tilde{\SF}(x,\hat{x}),\tilde{\rho}_u(\|\hat{u}\|_\infty ),\tilde{\varepsilon}\};
\end{align}
\end{enumerate}
\end{definition}
As demonstrated in \cite{swikir2019compositional}, the existence of an approximate alternating simulation function implies the existence of an approximate alternating relation from $T$ to $\hat{T}$. This relation ensures that for any behavior of $T$, there exists a corresponding behavior of $\hat{T}$ such that the distance between these behaviors is uniformly bounded by $\hat{\varepsilon} = \tilde{\alpha}^{-1}(\max{\tilde{\rho}_u(r), \tilde{\varepsilon}})$. For local abstraction, the concept of an $\varepsilon$-approximately alternating simulation function from $T_i$ to $\hat{T}_i$, for all $i \in \mathcal{N}$, is formally defined below.

\begin{definition}\label{def:SFD1}
Let $T_i=(X_i,X_{0_i},W_i,U_i,\TF_i,Y_i,\op_i)$  and $\hat T_i=(\hat X_i,\hat X_{0_i},\hat W_i,\hat U_i,\hat \TF_i,\hat Y_i,\hat \op_i)$ be transition subsystems with $\hat Y_i\subseteq Y_i$, $\hat \omega_i\subseteq W_i$. A function $ \mathcal{S}_i:X_i\times \hat X_i \to \mathbb{R}_{\geqslant0} $ is called a local alternating simulation function from $\hat T_i$ to $T_i$
if there exist $\alpha_i, \rho_{\omega_i}\in \mathcal{K}_{\infty}$,  $0<\sigma_i< 1$, $ \rho_{u_i} \in \mathcal{K}_{\infty}\cup \{0\} $, a symmetric matrix $D_i$ of appropriate dimension with conformal block partitions $D^{k j}_i, k, j \in\{1, 2\}$, and some $\varepsilon_i\in \mathbb{R}_{\geqslant 0}$ so that the following hold:
\begin{enumerate}
\item For every $ x_i\in X_i,\hat{x}_i\in\hat{X}_i$, we have, 
\begin{align}\label{lsf1}
	\alpha_i (\|x_i-\hat{x}_i\| ) \!\leqslant\! \SF_i(x_i,\hat{x}_i);
\end{align}
\item For every $x_i \!\in\!  X_i, \hat x_i \!\in\! \hat{X}_i,\hat{u}_i \!\in\!\hat{U}_i$ there exists $ u_i \!\in\! U_i$ such that for every $ \omega_i\in W_i, \hat \omega_i\in\hat{W}_i, x^+_i \!\in\! \TF_i(x_i,\omega_i,u_i)$ there exists $\hat{x}^+_i \!\in\! \hat{\TF}_i(\hat{x}_i,\hat{\omega}_i,\hat{u}_i)$ so that,
\begin{align}\label{lsf2}
	\SF_i(x^+_i,\hat{x}^+_i)\leqslant &		 \bar{\sigma}_i \SF_i(x_i,\hat{x}_i)+ \left[\begin{array}{c}\omega_i-\!\hat \omega_i \\ x_i-\hat{x}_i\end{array}\right]^{T} \underbrace{\left[\begin{array}{ll}D^{11}_i & D^{12}_i \\ D^{21}_i & D^{22}_i\end{array}\right]}_{D_i}\left[\begin{array}{c}\omega_i-\!\hat \omega_i \\ x_i-\hat{x}_i\end{array}\right]+\bar\rho_u(\|\hat{u}_i\|_{\infty} )+\bar\varepsilon_i.
\end{align}

\end{enumerate}
\end{definition}

The goal is to construct alternating simulation functions for the compound transition systems $T=\mathcal{I}(T_1,\ldots,T_N)$ and $\hat T=\mathcal{I}(\hat T_1,\ldots,\hat T_N)$ from the local alternating simulation functions of the subsystems. 
To achieve this goal, the following lemmas are recalled.

\section{Main Result}
The following theorem presents a compositional method for constructing an alternating simulation function from $\hat{T} = (\hat{T}_1, \ldots, \hat{T}_N)$ to $T = (T_1, \ldots, T_N)$ by utilizing local alternating simulation functions from $\hat{T}_i$ to $T_i$ for each $i \in \mathcal{N}$.

\begin{theorem}\label{thm:3}
Consider the interconnected transition system $T=\mathcal{I}(T_1,\ldots,T_N)$. Assume that each $T_i$ and its abstraction $\hat{T}_i$ admit a local alternating simulation function $\SF_i$ as in Defintion \ref{def:SFD1}.
If there exist $\mu_{i} \geq 0, i \in\{1 \ldots N\}$ such that 
		\begin{gather}\label{eq:compositional_constraint}
		{\left[\begin{array}{c}
		M \\
		I_{n}
		\end{array}\right]^{T} \mathbf{D}\left(\mu_{1} D_{1}, \ldots, \mu_{N} D_{N}\right)\left[\begin{array}{c}
		M \\
		I_{n}
		\end{array}\right] \preceq 0,} \\
        M \prod_{i=1}^{N} \hat{X}_{i} \subseteq \prod_{i=1}^{N} \hat{W}_{i},
		\end{gather}
        
		where
		\begin{align}
		\notag & \mathbf{D}\left(\mu_{1} D_{1}, \ldots, \mu_{N} D_{N}\right):= 
		 {\left[\begin{array}{llllll}
		\mu_{1} D_{1}^{11} & & & \mu_{1} D_{1}^{12} & & \\
		& \ddots & & & \ddots & \\
		& & \mu_{N} D_{N}^{11} & & & \mu_{N} D_{N}^{12} \\
		\mu_{1} D_{1}^{21} & & & \mu_{1} D_{1}^{22} & & \\
		& \ddots & & & \ddots & \\
		& & \mu_{N} D_{N}^{21} & & & \mu_{N} D_{N}^{22}
		\end{array}\right].} 
		\end{align}
Then, function $\tilde{\SF}:X\times \hat{X}\rightarrow \R_{\ge0}$ defined as,
\begin{align}\label{OF}
\tilde{\SF}&(x,\hat{x})\coloneq \sum_{i=1}^N \mu_{i} \SF_i(x_{i},\hat{x}_{i})  
\end{align}
is an alternating simulation function from $T=\mathcal{I}(T_1,\ldots,T_N)$ to  $\hat T=\mathcal{I}(\hat T_1,\ldots,\hat T_N)$. 
\end{theorem}

\bigskip

\section{Construction of Finite Abstractions}\label{1:IV}

In the previous section, we showed how to construct an abstraction of a system from the abstractions of its subsystems. In this section, our focus is on constructing a finite abstraction for an impulsive subsystem using an approximate alternating simulation. To ease readability, in the sequel, the index $i\in \n $ is omitted.

Consider an impulsive subsystem $\Sigma=(\R^{n},\mathbb W,\mathsf{W},\mathbb U,\mathsf{U}_{\tau},f,g)$, as defined in Definition \ref{def:sys1}. We restrict our attention to sampled-data impulsive systems, where the input curves belong to $\mathsf{U}_{\tau}$ containing only curves of constant duration $\tau$, i.e.,
\begin{align}\label{input}
\mathsf{U}_{\tau}=\{\nu:\R_{\ge0}\rightarrow \mathbb U| \nu(t)&=\nu((k-1)\tau), t\in [(k-1)\tau,k\tau),k\in\N_{\geqslant1}\}.
\end{align}
Moreover, we assume that there exist constant $\varphi$ such that for all $\omega\in \mathsf{W}$ the following holds,
\begin{align}\label{gb}
\|\omega(t)-\omega((k-1)\tau)\|\leqslant\varphi,\forall t\in [(k-1)\tau,k\tau),k\in\N_{\geqslant1}.
\end{align}

Next, we define sampled-data impulsive systems as transition subsystems. Such transition subsystems would be the bridge that relates impulsive systems to their finite abstractions.
\begin{definition}\label{tsm} Given an impulsive system $\Sigma=(\R^{n},\mathbb W,\mathsf{W},\mathbb U,\mathsf{U}_{\tau},f,g)$, we define the associated transition system $T_{\tau}(\Sigma)=(X,X_{0}, W,\mathcal{W},U,\mathcal{U},\TF)$ 
where:
\begin{itemize}
\item  $X=\R^{n} \times \{0,\ldots,\overline{z}\}$;
\item  $X_{0}=\R^{n}\times \{0\}$; 
\item $U=\mathbb U$;
\item $\mathcal{U}=\mathsf{U}_{\tau}$;
\item $W=\mathbb W$;
\item $\mathcal{W}=\mathsf{W}$;
\item  $(x^+,c^+)\in \TF((x,c),\omega,u)$ if and only if one of the following scenarios hold:
\begin{itemize} 
\item Flow scenario: $0\leq c\leq \overline{z}-1$, $x^+= \mathbf x_{x,\omega,u}(\pre{-}\tau)$, and $c^+=c+1$;
\item Jump scenario: $\underline{z}\leq c\leq \overline{z}$,  $x^+= g(x,\omega(0),u(0))$, and $c^+=0$. 
\end{itemize}
\end{itemize}
\end{definition}
For later use, define $\mathcal{W}_{\tau}$ as,
\begin{align}\label{ininput}
\mathcal{W}_{\tau}=\{\omega:\R_{\ge0}\rightarrow W| \omega(t)&=\omega((k-1)\tau),t\in [(k-1)\tau,k\tau),k\in\N_{\geqslant1}\}.
\end{align}
In order to construct a finite abstraction for $T_{\tau}(\Sigma)$, we introduce the following assumptions and lemmas. 

\begin{assumption}\label{likeiss}
Consider impulsive system $\Sigma=(\R^{n},\mathbb W,\mathsf{W},\mathbb U,\mathsf{U}_{\tau},f,g)$. Assume that there exist a locally Lipschitz function $ V:\R^{n}\times \R^{n} \to \R_{\geqslant0} $, $\mathcal{K}_{\infty}$ functions $\underline{\alpha}, \overline{\alpha}, \rho_{u_{c}},\rho_{u_{d}}$, and constants $\kappa_{c}\in\R,\kappa_{d}\in\R$, symmetric matrices $D_c$ and $D_d$ of appropriate dimensions with conformal block partitions $D^{k j}_m, k, j \in\{1, 2\}, m\in\{c, d\}$ such that the following hold,
\begin{itemize}
\item $\forall x,\hat x\in \R^{n}$,
\begin{align}\label{c1}
\underline{\alpha} (\| x-\hat{x}\| ) \leqslant V(x,\hat{x})\leqslant \overline{\alpha} (\| x-\hat{x}\|);
\end{align}
\item $\forall x,\hat x\in \R^{n}~\text{a}.\text{e}$, $\forall \omega,\hat{\omega}\in W$, and $\forall u,\hat u\in \mathbb{U}$,
\begin{align}\label{c2}
&\!\!\!\!\!\!\dfrac{\partial V(x,\hat{x})}{\partial x} \!f(x,\omega,u)\!+\!\dfrac{\partial V(x,\hat{x})}{\partial \hat{x}} \!f(\hat x,\hat{\omega},\hat{u})\\\notag
\!&\leqslant\! -\kappa_{c} V(x,\hat{x})\!+\left[\begin{array}{c}\omega-\!\hat \omega \\ x-\hat{x}\end{array}\right]^{T} \underbrace{\left[\begin{array}{ll}D^{11}_c & D^{12}_c \\ D^{21}_c & D^{22}_c\end{array}\right]}_{D_c}\left[\begin{array}{c}\omega-\!\hat \omega\\ x-\hat{x}\end{array}\right]+\!\rho_{u_{c}}(\| u\!-\! \hat{u}\|);
\end{align}\!\!\!\!\!
\item $\forall x,\hat x\in \R^{n}$,$\forall \omega,\hat{\omega}\in W$, and $\forall u,\hat u\in \mathbb{U}$,
\begin{align}\label{c3}
&\!\!\!\!\!V(g(x,\omega,u),g(\hat x,\hat{\omega},\hat{u}))\\\notag
\!&\leqslant\! \kappa_{d} V(x,\hat{x})+\left[\begin{array}{c}\omega-\!\hat \omega \\ x-\hat{x}\end{array}\right]^{T} \underbrace{\left[\begin{array}{ll}D^{11}_d & D^{12}_d \\ D^{21}_d & D^{22}_d\end{array}\right]}_{D_d}\left[\begin{array}{c}\omega-\!\hat \omega\\ x-\hat{x}\end{array}\right]+\rho_{u_{d}}(\| u \! -\! \hat{u}\| ).
\end{align}
\end{itemize}  
\end{assumption}
\begin{assumption}\label{ass2} 
There exist $\mathcal{K}_{\infty}$ function $\hat{\gamma}$ such that for all $x,y,z \in \R^{n}$,
\begin{align}\label{tinq} 
V(x,y)\leqslant V(x,z)+\hat{\gamma}(\| y-z\|).
\end{align}
\end{assumption}

We now have all the ingredients to construct a finite abstractions $\hat{T}_{\tau}(\Sigma)$ of transition system $T_{\tau}(\Sigma)$ associated with the impulsive system $\Sigma$ admitting a function $V$ that satisfies Assumption \ref{likeiss} as follows. 
\begin{definition}\label{smm} Consider a transition system $T_{\tau}(\Sigma)=(X,X_{0}, W,\mathcal{W},U,\mathcal{U},\TF)$, associated to the impulsive system $\Sigma=(\R^{n},W,\mathsf{W},\mathbb U,\mathsf{U}_{\tau},f,g)$. Assume $\Sigma$ admits a function $V$ that satisfies Assumption \ref{likeiss}. One can construct finite abstraction $\hat T_{\tau}(\Sigma)=(\hat{X},\hat{X}_{0},\hat{W},\hat{\mathcal W},\hat{U},\hat{\mathcal U},\hat{\TF})$ where:
\begin{itemize}
\item  $\hat X=\hat{\R}^{n} \times \{0,\ldots,\overline{z}\}$, where $\hat{\R}^{n}=[\R^{n}]_{\eta^x}$ and $\eta^x$ is the state set quantization parameter;
\item  $\hat X_{0}=\hat{X}\times \{0\}$; 
\item $\hat{W}=[W]_{\eta^\omega}$, where $\eta^\omega$ is the internal input set quantization parameter;
\item $\hat{\mathcal W}=\{\hat{\omega}:[0,\tau]\rightarrow \hat{W}| \hat{\omega}\in{\mathcal W}_{\tau}|_{[0,\tau]}\}$;
\item $\hat{U}=[U]_{\eta^u}$, where $\eta^u$ is the external input set quantization parameter;
\item $\hat{\mathcal U}=\{\hat{u}:[0,\tau]\rightarrow \hat{U}| \hat{u}\in{\mathcal U}|_{[0,\tau]}\}$;
\item  $(\hat x^+,c^+)\in \hat\TF((\hat x,c),\hat{\omega},\hat u)$ iff one of the following scenarios hold:
\begin{itemize} 
\item Flow scenario: $0\leqslant c\leqslant \overline{z}-1$, $|\hat x^+-\mathbf{x}_{\hat{x},\hat{\omega},\hat{\nu}}(\tau)|\leqslant\eta^x$, and $c^+=c+1$;
\item Jump scenario: $\underline{z}\leqslant c\leqslant \overline{z}$,  $|\hat x^+ - g(\hat x,\hat{\omega}(0),\hat u(0))|\leqslant\eta^x$, and $c^+=0$. 
\end{itemize}
\end{itemize}
\end{definition}
 
In the definition of the transition function, and throughout the remainder of this paper, we adopt a slight abuse of notation by identifying $\hat{u}$ (respectively, $\hat{\omega}$) with the constant external (respectively, internal) input curve defined on the domain $[0, \tau)$ and taking the value $\hat{u}$ (respectively, $\hat{\omega}$).

Now, we establish the relation from $T_{\tau}(\Sigma)$ to $\hat T_{\tau}(\Sigma)$, introduced above, via the notion of alternating simulation function as in Definition \ref{sf}.
\begin{theorem}\label{thm1}
Consider an impulsive system $\Sigma=(\R^{n},W,\mathsf{W},\mathbb U,\mathsf{U},f,g,)$ with its associated transition system $T_{\tau}(\Sigma)=(X,X_{0}, W,\mathcal{W},U,\mathcal{U},\TF)$. Suppose Assumptions \ref{likeiss}, and \ref{ass2} hold. Consider finite abstraction $\hat T_{\tau}(\Sigma)=(\hat{X},\hat{X}_{0},\hat{\omega},\hat{\mathcal W},\hat{U},\hat{\mathcal U},\hat{\TF})$ constructed as in Definition \ref{smm}. If inequality,
\begin{align}\label{cc1}
\ln(\kappa_{d})-\kappa_{c}\tau c<0,
\end{align}
holds for $c\in\{\underline{z},\overline{z}\}$, then function $\mathcal{V}$ defined as,
\begin{align}\label{sm}
\mathcal{V}((x,c),(\hat{x},c))\!\!\Let\!\!\left\{
\begin{array}{lr} 
\!\!\!\!V(x,\hat{x})    \quad\,~ \text{if}\quad \kappa_{d}<1 \!~\&~ \kappa_{c}>0,\\
\!\!\!\!\dfrac{V(x,\hat{x})}{e^{-\kappa_{c}\tau \epsilon c}}   \quad \text{if}\quad \kappa_{d}\geqslant1 ~\&~ \kappa_{c}>0,\\
\!\!\!\!\dfrac{V(x,\hat{x})}{\kappa_{d}^{-\frac{c}{\delta}}} \quad\, \text{if}\quad \kappa_{d}<1 ~\&~ \kappa_{c}\leqslant0,
\end{array}\right.
\end{align}
for some $0<\epsilon<1$ and $\delta>\overline{z}$, is an alternating simulation function from $\hat T_{\tau}(\Sigma)$ to $T_{\tau}(\Sigma)$. 
\end{theorem}

\bibliographystyle{ieeetr}      
\bibliography{arxive}

\end{document}